\documentclass{article}

\usepackage{amssymb,amsmath,xcolor,graphicx,xspace,colortbl,rotating} %
\usepackage[raggedrightboxes]{ragged2e} 
\usepackage{amsfonts}  
\usepackage{amssymb}  
\usepackage{float}  
\usepackage{graphics}  
\usepackage{ragged2e}  
\usepackage{wrapfig}  
\usepackage{xcolor}  
\graphicspath{{QUECST20_graphics/}{QUECST20_tcache/}{QUECST20_gcache/}}
\DeclareGraphicsExtensions{.pdf,.eps,.ps,.png,.jpg,.jpeg}

  \input{tcilatex} 
\begin{document}

\title{Auto-stabilized Electron
}
\author{Munawar Karim
\\\relax
Department of Physics, St. John Fisher College, Rochester, NY 14618
}
\date{ April, 2019
}
\maketitle
\begin{abstract}We include effects of self-gravitation in the self-interaction of single electrons
with the electromagnetic field. When the effect of gravitation is included there is an inevitable
cut-off of the k-vector - the upper limit is finite. The inward pressure of the self-gravitating
field balances the outward pressure of self-interaction. Both pressures are generated by
self-interactions of the electron with two fields - the vacuum electromagnetic
field and the self-induced gravitational field. Specifically we demonstrate that gravitational
effects must be included to stabilize the electron. We use the
Einstein equation to perform an exact calculation of the bare mass and electron radius. We find a
close-form solution. We find the electron radius
$r_{e} =9.2\sqrt{\alpha / 4\pi } \sqrt{\hbar  G/c^{3}}$
$ =9.2\sqrt{\alpha / 4\pi }l_{P}$
$ \approx 10^{ -36}m$
.
$\sqrt{\hbar  G/c^{3\text{}}}$
is the Planck length
$\ell _{P}$
, which is educed from first principles. We find that the electromagnetic and gravitational
fields merge at
$\left (8/3\right )\sqrt{\alpha / 4\pi } \sqrt{\hbar  c/G}$
$ =\left (8/3\right )\sqrt{\alpha / 4\pi }\ m_{P} =10^{17} G e V$
in terms of the Planck mass
$m_{P}$
. Renormalisation is accomplished by requiring continuity of the interior and exterior
metrics at
$r_{e}$
.

PACS 04.62.+v, 11.10.Gh, 12.10.-g, 12.20.Ds
\end{abstract}
\begin{center}
\bigskip \ I. INTRODUCTION
\end{center}\par
\bigskip The paradox of electron stability has been recognized since its discovery \cite{JJ}. Instability is inevitable in any system with charge distributed
over an extended volume. There have been several conjectures to explain stability (i) \ a shell of non-electromagnetic origin to contain the field
(Poincare' stress), (ii) using radiation reaction to compensate the outward pressure
(Abraham-Lorentz equation), (iii) dimensional regularization etc.

Poincare's suggestion was to trap the charge in a rigid shell subjecting it to a
stress \cite{Poincare}. This was found to be untenable.

When radiation reaction is included as in the Abraham-Lorentz equation \cite{abraham}, one gets runaway solutions instead of stability.

Yet another suggestion was to treat the space dimension as a free parameter \textit{d
}(dimensional regularization) \cite{bolini}, then use its
variation to solve a divergent integral. Choosing \textit{d}=4 unfortunately leads to a
pole, which then must be compensated, raising new issues.

It was recognized that renormalisation was required to compute the self-energy of the
electron. In this scheme the measured mass of electrons
$m_{\;\text{}\;}$
was used to terminate the divergence.

With the advent of quantum electrodynamics new schemes were proposed, but
even these were partially successful. Although the calculations are well-established, the result is
inevitably divergent (albeit logarithmically at best).
Various work-arounds have been proposed, among them a modification of electrodynamics at short
distances, terminating the integral at the Compton or Planck
length etc. We will discuss some of these later.

These approaches have one common motivation: \ that
infinities are a mathematical anomaly which must be side-stepped, removed, truncated or at worst
ignored. Needless to say these approaches suffer from deficiencies.
Infinities remain because they are intrinsic to the theory.

We claim instead that the appearance of infinities is the consequence
of ignoring basic physical phenomena. The infinities are real. In the sections below we describe
what these are and calculate the resulting corrected mass from first principles.

\begin{center}
A. Computation algorithm
\end{center}\par
Referring to a free electron, Feynman's final result is
\begin{equation} \Delta m =\frac{4 \pi  e^{2}}{2 m i} \int _{ -\infty }^{\infty }\frac{\tilde{u} \left (2 m +2 ^{\not}k\right ) u}{k^{2} -2 \vec{p} \cdot \vec{k}} \frac{d^{4}k}{\left (2 \pi \right )^{4}} \frac{1}{k^{2}} \label{Feynman}
\end{equation}
for the mass correction
$ \Delta m$
\cite{Feynman} where
$m$
is the experimental mass and
$e$
the charge of the electron ;
$k$
is the four-momentum of photons and
$p$
is the four-momentum of the electron.

The integral is of the form
$d^{4} k/k^{4}$
which is intrinsically divergent. Feynman gets around the divergence by modifying the photon
kernel to
$\left (1/k^{2}\right ) c \left (k^{2}\right )$
from
$\left (1/k^{2}\right )\text{.}$
The function
$c \left (k^{2}\right )$
is chosen so that
$c \left (0\right ) =1$
and
$c \left (k^{2}\right ) \rightarrow 0$
as
$k^{2} \rightarrow \infty $
. This conjectured change in the photon kernel is a consequence of altering the laws of
electrodynamics at short distances. With this alteration the integral is still divergent, albeit
logarithmically; a free parameter is introduced. As a consequence the high frequency components
of the Fourier expansion or equivalently, the short-range contributions are modified. No reason
exists to justify this alteration, since there is no evidence
that the laws of electrodynamics are inadequate at short distances. The puzzle remains unsolved.

\begin{figure}\centering 
\setlength\fboxrule{0in}\setlength\fboxsep{0.1in}\fcolorbox[HTML]{FFFFFF}{FFFFFF}{\includegraphics[ width=2.59375in, height=1.0520833333333335in,]{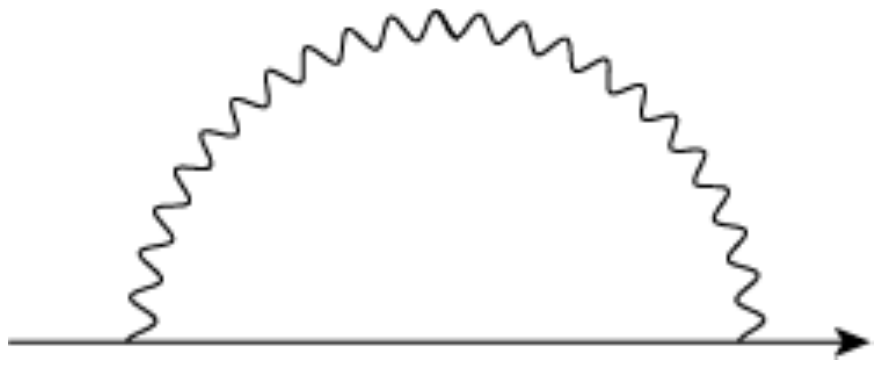}
}
\caption{\ Emission and absorption of single
photon
}\label{Fig.1}\end{figure}

The diagram Fig. \ref{Fig.1} represents the
electron emitting and absorbing a single photon

\begin{center}
\textit{1.}\textit{Near field}
\end{center}\par
A way out of this dilemma is to re-interpret
the one-loop correction and relate it to fields in the vicinity of a classical radiation source.
Surrounding any source radiating at a wavelength
$\lambda $
there is a near zone which extends to
$r =\lambda /2 \pi \text{.}$
In terms of the wave-vector
$\kappa $
the condition that makes this so is
$\kappa  \cdot r =1$
. The extent of the near zone scales as the inverse of the wave-vector. Within the near zone
the fields are Coulomb-like. \ The fields exert a radially
outward pressure. Outside the near zone, fields acquire a transverse component; they propagate.

Energy is stored within the near zone. It is this energy that shows up as an excess
mass in Eq.(\ref{Feynman}). Energy is continuously exchanged
between fields and source in the near zone ; pictorially represented as the emission and absorption
of single photons Fig.(\ref{Fig.1}). Higher order diagrams
correspond to quadrupole, octupole and higher order fields.
Appropriately, fields in this zone may be treated as that in the diagram above.

\begin{center}
\bigskip B. Gravitation
\end{center}\par
We can see how gravity comes into the picture.
If we squeeze an electron into a sphere of radius equal to a Compton wavelength (
$\lambda _{C}$
) the energy density can be estimated. Setting
$r =\lambda _{C}$
\begin{equation}\frac{e^{2}}{4 \pi  \epsilon _{0} r} \frac{1}{\left (4 \pi /3\right ) r^{3}} =\frac{e^{2}}{4 \pi  \epsilon _{0}} \frac{3}{4 \pi } \frac{1}{\left (\lambda _{C}\right )^{4}} =\frac{e^{2}}{4 \pi  \epsilon _{0}} \frac{3}{4 \pi } \genfrac{(}{)}{}{}{m c}{h}^{4} =10^{18}\; J/m^{3} \label{unitP}
\end{equation}
which is
$ \approx 10^{13}$
atmospheres. By comparison the pressure at the center of the Sun is
$10^{11}$
atmospheres. This shows that such high energy densities are not just the purview of
astrophysical sources but are common among elementary particles.

Clearly under these conditions energy densities are high enough to alter the metric in
the vicinity of the electron. At these densities virtual excitations follow geodesics of curved
metrics rather than flat space. Virtual excitations loop back to the source along geodesics of the
distorted metric. For example in Fig.(\ref{Fig.1}) the emission
and absorption of virtual photons occurs in a curved
metric. General relativistic effects not only cannot be ignored; they become essential part of the
dynamics of the electron. Gravitation is part of the electron.

It is evident that theories that rely on flat space geometry are inadequate; the
engendered divergences are evidence that in such theories a major reservoir of energy is being
ignored. The enormous outward forces cannot be balanced in flat space; curved space-time must be
included.

There have been attempts to include gravity in QED phenomena. An early example is
Isham, Salam and Strathdee, \cite{salam} and others.

\begin{center}
\textit{1. Gravitating electron}
\end{center}\par
In this paper we insert gravitation first by integrating
Eq.(\ref{Feynman}) up to an upper limit for
$k$
. The upper limit is an unknown for now. We set the momentum
\begin{equation}k =\hbar  \kappa  =\hbar  \frac{2 \pi }{\lambda } \label{vector}
\end{equation}
where
$\kappa $
is the wave number. The corresponding near zone radius for
$\lambda $
is
$r =\lambda /2 \pi \text{.}$

Performing the integral Eq.( \ref{Feynman}) we
get for the mass correction (see Appendix)
\begin{equation} \Delta m \equiv \mu  \left (\eta \right ) =\frac{\alpha  m}{2 \pi } \left [ -\frac{\eta }{2} \sqrt{1 +\frac{1}{\eta }} +\eta  +\ln  \left \{\sqrt{\eta } \left (\eta  \sqrt{1 +\frac{1}{\eta }} +1\right )\right \}\right ] \label{mass}
\end{equation}
in terms of a dimensionless variable
\begin{equation}\eta  \equiv \hbar /2 m c r =\lambda _{C}/4 \pi  r
\end{equation}

We have redefined
$ \Delta m \equiv \mu  \left (\eta \right )$
.

There is an energy density associated with the self-field within the near zone. The
energy density, or the stress tensor, alters the metric within the near zone. The net result is an
inward pressure. Analogous to the Sun where the radiation pressure (or Fermi pressure
in the case of white dwarfs or neutron stars) is balanced by the inward gravitational pressure.

It is this pressure that compensates
the outward pressure from the electron field that stabilises the electron.

Increasing values of the
$k -$
vector raise the stress tensor which in turn increases the inward gravitation induced
pressure. The electron is auto-stabilized.

Equating the two competing pressures yields an upper limit for
$k\text{.}$
We will calculate this limit.

We use the Einstein equation to calculate the resulting Einstein tensor.
We start with a line element of the form
\begin{equation}d s^{2} = -e^{2 \Phi } c^{2} d t^{2} +e^{2 \Lambda } d r^{2} +r^{2} \left (d \theta ^{2} +\sin ^{2} \theta \; d \phi ^{2}\right )
\end{equation}
Fields within the near zone are treated as a perfect fluid. The elements of the stress tensor
are
\begin{equation}T^{\widehat{0} \widehat{0}} =\rho  ,\text{}T^{\widehat{r} \widehat{r}} =T^{\widehat{\theta } \widehat{\theta }} =T^{\widehat{\phi } \widehat{\phi }} =P
\end{equation}
in the fluid's orthonormal rest-frame basis vectors. Imposing momentum conservation and
spherical symmetry the relevant Einstein equations are
\begin{equation}G^{\widehat{0} \widehat{0}} = -\frac{1}{r^{2}} \frac{d}{d r} \left [r \left (1 -e^{ -2 \Lambda }\right )\right ] =8 \pi  T^{\widehat{0} \widehat{0}} =8 \pi  \rho  \label{grav1}
\end{equation}
\begin{equation}G^{\widehat{1} \widehat{1}} = -\frac{1}{r^{2}} \frac{d}{d r} \left (1 -e^{ -2 \Lambda }\right ) +\frac{2}{r} e^{ -2 \Lambda } \frac{d \Phi }{d r} =8 \pi  T^{\widehat{1} \widehat{1}} =8 \pi  P \label{grav}
\end{equation}
We define a new metric coefficient
$\mu  \left (r\right )$
(same as in Eq.(\ref{mass}) as
\begin{equation}g_{1 1} =e^{2 \Lambda } \equiv \frac{1}{1 -\frac{2 \mu }{r}} \label{renorm}
\end{equation}
$\mu  \left (r\right )$
is the corrected mass inside the radius
$r$
. The time-time component of Eq.(\ref{grav1}) takes the
form
\begin{equation}\frac{d \mu }{d r} =4 \pi  r^{2} \rho 
\end{equation}
whereas radial-radial component of Eq. (\ref{grav} ) takes
the form
\begin{equation}\frac{d \Phi }{d r} =\frac{\mu  +4 \pi  r^{3} P}{r \left (r -2 \mu \right )} \label{phi1}
\end{equation}
The proper density is
\begin{equation}\rho  =\frac{1}{4 \pi  r^{2}} \frac{1}{\sqrt{\left \vert g_{11}\right \vert }} \frac{d \mu }{d r} \label{density}
\end{equation}
Since the volume element
\begin{equation}d V =\sqrt{\vert g_{1 1}\vert } r^{2} \sin  \theta \; d \theta \; d \phi \; d r
\end{equation}
These equations when combined with the condition for hydrostatic equilibrium lead to the
Tolman-Oppenheimer-Volkov equation
\begin{equation} -\frac{d P \left (r\right )}{d r} =\left (\rho  \left (r\right ) +P \left (r\right )\right ) \genfrac{(}{)}{}{}{\mu  \left (r\right ) +4 \pi  r^{3} P \left (r\right )}{r^{2} \left (1 -\frac{2 \mu  \left (r\right )}{r}\right )} \label{TOV}
\end{equation}
This first order differential equation can be solved once the relation between pressure
$P (r)$
and density
$\rho  \left (r\right )$
is known. The negative slope guarantees the decrease of
$P (r)$
until for some value of
$r$
,
$P (r) =0$
. We seek this value of
$r$
.

\begin{center}
C. Equation of state
\end{center}\par
We can derive an equation of state - relate
$P \left (r\right )$
with
$\rho  \left (r\right )\text{.}$
We use the work equation
\begin{equation}d E = -P\; d V ;\text{}\frac{d E}{d V} = -P
\end{equation}
Also
$E =\rho  V\text{}$
\begin{eqnarray}d E &  = & V\; d \rho  +\rho \; d V \\
\frac{d E}{d V} &  = & \frac{d \rho }{d V} V +\rho  = -P \\
\frac{d \rho }{d V} &  = & \frac{d \rho }{d r} \frac{d r}{d V} =\frac{d \rho }{d r} \genfrac{(}{)}{}{}{3}{4 \pi }^{1/3} V^{ -2/3} \\
\frac{d E}{d V} &  = & \frac{d \rho }{d r} \genfrac{(}{)}{}{}{3}{4 \pi }^{1/3} V^{ -2/3} V +\rho  = -P\end{eqnarray}
The near zone fields have an equation of state which is

\begin{equation}P = -\rho  -r \frac{d \rho }{d r}
\end{equation}
Or in terms of
$\mu  \left (r\right )$
the equation of state can be written as
\begin{equation}P =\frac{1}{4 \pi  r^{2}} \frac{1}{\sqrt{\left \vert g_{11}\right \vert }} \frac{d \mu }{d r} -\frac{1}{4 \pi  r} \frac{1}{\sqrt{\left \vert g_{11}\right \vert }} \frac{d^{2} \mu }{d r^{2}}
\end{equation}
Eq.(\ref{TOV}) can be re-written in terms of the variable
$\eta \text{.}$
\begin{equation*}\frac{d P}{d \eta } =\frac{d P}{d r} \frac{d r}{d \eta }
\end{equation*}
We calculate
$P \left (\eta \right )$
by integrating
$\frac{d P \left (\eta \right )}{d \eta }$
\begin{equation}P \left (\eta \right ) =\int \frac{d P \left (\eta \right )}{d \eta } d \eta  \label{press}
\end{equation}
and find the root of the integral
$\eta _{e}$
for which
$P \left (\eta _{e}\right ) =0$
; this is the value of
$\eta $
and thus the lower limit
$r$
or the upper limit
$k$
we are looking for.

\begin{center}
D. Results
\end{center}\par
\ In order to simplify the computation we will use an
approximation where
$\eta  \gg 1$
to find the root of Eq.(\ref{TOV})
. This allows us to simplify the terms
$\mu  \left (\eta \right )$
,
$\rho  \left (\eta \right )$
and
$P (\eta )$
(which has
$\frac{ \partial \rho }{ \partial \eta }$
).

For example for
$\eta  \gg 1$
,
$\rho  \left (\eta \right ) \rightarrow \frac{\eta ^{4}}{2} ,\frac{ \partial \rho }{ \partial \eta } \rightarrow 2 \eta ^{3}\text{,}$
$\mu  \left (\eta \right ) \rightarrow \frac{\eta }{2}$
.  For
$\eta $
we get a fourth order algebraic equation with roots, of which one is positive, one negative
and two complex. We take the positive root and get

\begin{equation}\eta _{e} =\frac{8}{15}\sqrt{2}\left (\frac{c^{2}}{2mG}\frac{2\pi }{\alpha }\frac{\hbar }{2mc}\right )^{1/2} =\frac{8}{15}\frac{1}{m}\sqrt{\frac{\pi }{\alpha }}\sqrt{\frac{\hbar c}{G}} \label{eta0}
\end{equation}
where
$P \left (\eta \right ) =0$
.

 The corresponding electron radius is

\begin{equation}r_{e} =7.5\sqrt{\frac{\alpha }{4\pi }}\sqrt{\frac{\hbar G}{c^{3}}}
\end{equation}
The electron radius is greater than the Schwarzschild radius
$2m$
. We note that the interior line element develops a singularity in
$g_{11}$
when
\begin{equation}1 -\frac{\alpha }{\pi } \frac{m G}{c^{2}} \frac{2 m c}{\hbar } \frac{\eta ^{2}}{2} =0
\end{equation}
That is when
\begin{equation}r_{0} =\sqrt{\frac{\alpha }{4\pi }\frac{\hbar G}{c^{3}}}
\end{equation}
also
$r_{0} >2m .$
And so the electron radius
\begin{equation}r_{e} =7.5r_{0} =2.95 \times 10^{-36}m\
\label{re}
\end{equation}
We observe that
$r_{e} >r_{0}$
- the pressure falls to zero outside the singularity.

Simplifying the equation the radius is
\begin{equation}r_{e} =7.5\sqrt{\frac{G}{c^{4}}\frac{e^{2}}{\pi }} \label{Planck}
\end{equation}
The radius is independent of the electron mass
$m$
and
$\hbar $
and is entirely in terms of fundamental constants
$e$
,
$G$
and
$c$
. Since it is a radius independent of the electron mass we re-write the radius as a universal
radius
$r_{ \ast }$
in terms of the Planck length as
\begin{equation}r_{ \ast } =7.5\sqrt{\frac{\alpha }{4\pi }}l_{P} \label{rr}
\end{equation}
We note that what is called the Planck length
$\ell _{P}$
appears naturally.

 If instead we integrate
Eq.(\ref{press}) numerically we find
$P \left (\eta \right ) =0$
when
\begin{equation}r_{e} =9.21\sqrt{\frac{\alpha }{4\pi }\frac{\hbar G}{c^{3}}} =5.5 \times 10^{ -36} m
\end{equation}

or
\begin{equation}r_{e} =9.21\sqrt{\frac{\alpha }{4 \pi }} \ell _{P}
\end{equation}
which is close to Eq.(\ref{rr}).

One may ask if general relativity is valid at such short lengths. The existence of
black holes and the Big Bang offer proof that there is no known lower limit to lengths in general
relativity.

In terms of energy
\begin{equation}r_{ \ast } =10^{20} G e V
\end{equation}
If we substitute
$\eta _{e}$
from Eq.(\ref{eta0}) into Eq.(\ref{mass}) we get a mass independent of the electron mass. We call it the universal mass
$\mu _{ \ast }$
\begin{equation}\mu _{ \ast } =\frac{8}{3}\sqrt{\frac{\alpha }{4\pi } \frac{\hbar  c}{ G}} =\frac{8}{3}\sqrt{\frac{\alpha }{ 4\pi }} m_{P} \label{Planck1}
\end{equation}
in terms of the Planck mass
$m_{P}\text{.}$
The Planck mass also appears naturally.

Simplifying the result
\begin{equation}\mu _{ \ast } =\frac{8}{3}\sqrt{\frac{1}{4\pi }}\sqrt{\frac{e^{2}}{  G}}
\end{equation}
dependent on
$e$
and
$G$
alone and       independent of
$\hbar $
and
$c$
. Numerically
\begin{equation}\mu _{ \ast } \approx 10^{17}\; G e V \label{bare}
\end{equation}

This result is remarkable for several reasons. Since
$\mu _{ \ast }$
is enormously larger than the physical mass
$m$
it cannot be the corrected mass. Furthermore
$\mu _{ \ast }$
is independent of mass; it is solely in terms of the fundamental constants
$e$
and
$G$
. Since it depends on
$e^{2}$
it is also independent of the sign of the charge. The inescapable conclusion is that the
result is a general result applicable to all charged particles irrespective of the mass and sign of
the charge.

Although our goal was to derive the corrected electron mass we have found instead a
mass that applies to all charged particles. A possible interpretation is that
$\mu _{ \ast }$
is the universal bare mass.

These observations also apply to the radius
$r_{e}$
since it too is independent of mass, and is in terms of the fundamental constants
$G ,$
$c$
and
$e$
.

The value of
$\mu _{ \ast \text{}}$
is close to the GUT energy
$\left (10^{16}\; G e V\right )$
where it is conjectured that all forces except gravity merge. It would appear that all
forces, including gravity, merge at
$10^{17}\; G e V$
.

Since the unified field energy is independent of
$\hbar $
one may also conclude that the unified field is not quantized but is a continuum. This is a
serendipitous result.

\

The bare mass, is now exact. The integral converges to an exact value.
Physical laws remain unaltered. The momentum upper limit is
\begin{equation}k^{\max } =\frac{\hbar }{r_{e}}\text{}
\end{equation}

We reiterate that this is a self-regulating mechanism since if
$k$
exceeds
$k^{\max }$
it engenders a proportional reaction from the metric such that the system reverts to a state
of equilibrium. The equilibrium is stable.

At
$r =10^{ -36} m$
where the two competing pressures are equal; the pressures are
\begin{equation}10^{70} m^{ -2}
\end{equation}
in geometric units, or
$10^{114}\; N/m^{2}$
in standard units. Evidently the electron surface is highly stressed; the pressure is
$ \approx $
$10^{109}$
atmospheres on the surface. By comparison the pressure inside neutron stars is merely
$ \approx $
$10^{30}$
atmospheres.

At this value the outward pressure due to the energy density of self-interaction
equals the inward pressure of the curved metric.

Pictorially, the distorted metric in the vicinity of the electron looks like this
photo-representation Fig.(\ref{Fig.2}):

\begin{figure}[t]\centering 
\setlength\fboxrule{0in}\setlength\fboxsep{0.1in}\fcolorbox[HTML]{FFFFFF}{FFFFFF}{\includegraphics[ width=3.8541666666666665in, height=2.9270833333333335in,]{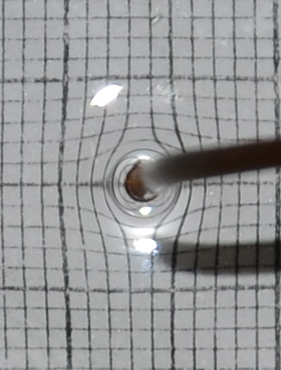}
}
\caption{\ Pictorial representation of metric
alteration due to self-gravity of self-energy of electron. The dark disk is a stand-in for the
electron. The photograph is that of a wire tip immersed in a thin layer of water in a clear glass
baking dish. A graph paper is glued to the underside of the dish. Surface tension distorts the water
surface
}\label{Fig.2}\end{figure}

The inward pressure is a consequence of the distorted metric.

\begin{center}
E. Renormalization
\end{center}\par
We are left to compute
$g_{0 0}$
from Eq.(\ref{phi1})
\begin{equation*}g_{0 0} =e^{2 \Phi }
\end{equation*}
where
\begin{equation}\frac{d \Phi _{i}}{d r} =\frac{\mu  +4 \pi  r^{3} P}{r^{2} (1 -\frac{2 \mu }{r})} \label{phi}
\end{equation}
We calculate the interior metric element
$\Phi _{i}$
by integrating Eq.(\ref{phi}). As expected
$\Phi _{i} =0$
on the surface defined by the radius
$r_{e}$
. Integration also yields a constant
$C$
. We choose the constant
$C$
to match the interior and exterior solutions at the surface. This is how we renormalize the
mass. At the surface we require
\begin{equation}\Phi _{e} =\Phi _{i}
\end{equation}
The exterior metric is the Schwarzschild solution for an electron of mass
$m$
. So for
$r >r_{e}$

\begin{equation*}g_{0 0} =e^{2 \Phi _{e}} =\left (1 -\frac{2 m}{r}\right )
\end{equation*}
and for
$r \leq r_{e}$
\begin{equation}g_{0 0} =\exp  \left (1 -\frac{2 m}{r_{e}}\right )\exp 2 \Phi _{i}
\end{equation}
where
$\Phi _{i}$
is obtained from Eq.(\ref{phi}). The same algorithm is used
to ensure the continuity of the
$g_{1 1}$
term (Eq.(\ref{renorm}).

Although we have renormalized the metric for electrons,
we note that for other particles such as protons or neutrons the same algorithm can be used with the
appropriate masses instead.

Although there have been previous attempts to introduce gravitation to remove
infinities in the self-energy correction to the electron mass \cite{salam} we believe this work demonstrates the existence of an auto-generated curved
metric in the vicinity of an electron as well as an explicit computation of
the upper limit on the momentum vector. The self-energy integral is shown to be finite; infinities
in the integral have been removed. The mass correction has an exact value, free of infinities.

We have demonstrated that gravitation is essential to stabilize the electron.
Competing pressures from the electromagnetic and gravitational fields auto-stabilize the electron.
We have identified the Poincare shell.

Furthermore instead of using dimensional analysis to derive the Planck length and mass
we show that both are a consequence of merging quantum electrodynamics with
gravitation. We calculate from first principles the radius of the electron as well as the Planck
length and Planck mass. In deriving the electron radius
and bare mass we identify the mechanism (inward pressure of the induced metric) that stabilises the
electron. The mass correction is close to the GUT scale
- an unexpected result. In demanding a smooth transition of the metric across the electron surface
we uncover a rigorous mechanism for renormalisation.

The solution is exact; since it is independent of mass it is valid for strong fields
as well.

\bigskip
\begin{center}
H. Acknowledgments
\end{center}\par
\begin{flushleft}We are grateful to Belal \ Baaquie who
identified the connection between Eq.(\ref{Planck}) and the
Planck length, to Ashfaque Bokhari for guidance through the labyrinth of general relativity, to
Roberto Onofrio for continuous encouragement and advice, to Adrian Mellissinos who identified
Eq.(\ref{bare}) as the bare mass; and for support from my wife Pat
and daughter Kim. This work was completed during a sabbatical leave as visiting professor at the
Namibia University of Science and Technology in Windhoek. We acknowledge the hospitality afforded by
Professor Tjama Tjivikua, Vice Chancellor of the university and Professor Habauka Kwaambwa, Head of
Department, Natural and Applied Sciences. An abbreviated version of this paper was published
(\cite{IJMA}).
\end{flushleft}\par
\begin{center}
I. Appendix
\end{center}\par
Derivation of Eq.(\ref{Feynman}):\ start by using the identity: we follow Feynman \cite{Feynman}.
\begin{equation}\frac{1}{a b} =\int _{0}^{1}\frac{d x}{\left [a x +b \left (1 -x\right )\right ]^{2}}
\end{equation}

Define two quantities
\begin{equation}k^{2} -2 p \cdot k \equiv a ;\text{}k^{2} \equiv b
\end{equation}

Then
\begin{equation}\frac{1}{\left (k^{2} -2 p \cdot k\right ) k^{2}} =\int _{0}^{1}\frac{d x}{\left [\left (k^{2} -2 p \cdot k\right ) x +k^{2} \left (1 -x\right )\right ]^{2}} =\int _{0}^{1}\frac{d x}{\left [k^{2} -2 p \cdot k x\right ]^{2}}
\end{equation}

So Eq.(\ref{Feynman})) becomes
\begin{equation}\mu  =\frac{4 \pi  e^{2}}{2 m i} \int _{ -\infty }^{\infty }\int _{0}^{1}\frac{\tilde{u} (2 m +2^{\not} k ) u}{\left [k^{2} -2 p \cdot k x\right ]^{2}} \frac{d^{4} k}{\left (2 \pi \right )^{4}} \frac{d x}{k^{2}}
\end{equation}

\begin{equation}d^{4} k =d \omega \; d^{3} \vec{K} ;\text{}k^{2} -2 p \cdot k x =\omega ^{2} -\left (\vec{K}^{2} +2 p \cdot k x\right )
\end{equation}

Let
$x p \equiv p$
then
\begin{equation}\mu  =\frac{4 \pi  e^{2}}{2 m i} \frac{1}{p} \int _{ -\infty }^{\infty }\int _{0}^{1}\frac{\tilde{u} (2 m +2^{\not} k ) u}{\left [\omega ^{2} -\left (\vec{K}^{2} +2 p \cdot k x\right )\right ]^{2}} \frac{d \omega \; d^{3} \vec{K}}{\left (2 \pi \right )^{4}} \frac{d x}{k^{2}}
\end{equation}

Do the
$\omega $
integral first using residues and for
$\varepsilon  \ll L +\vec{K}^{2}$
:
\begin{equation}\int _{ -\infty }^{\infty }\frac{d \omega }{\left [\omega ^{2} -i \varepsilon  +\left (\vec{K}^{2} +2 p \cdot k\right )\right ]} =\frac{ -2 \pi  i}{2 \sqrt{\vec{K}^{2} +2 p \cdot k}} \equiv \frac{ -2 \pi  i}{2 \sqrt{\vec{K}^{2} +L}}
\end{equation}

Since the only contribution to the integral comes from the pole we choose the limits
as
$ \pm \infty $
for
$\omega $
. Take derivatives with respect to
$L$
on both sides:
\begin{equation}\int _{ -\infty }^{\infty }\frac{d \omega }{\left [\omega ^{2} -i \varepsilon  +\left (\vec{K}^{2} +2 p \cdot k\right )\right ]^{2}} =\frac{\pi  i}{2 \left (\vec{K}^{2} +2 p \cdot k\right )^{3/2}}
\end{equation}

\begin{equation}\int _{0}^{\infty }\frac{d \omega }{\left [\omega ^{2} -i \varepsilon  +\left (\vec{K}^{2} +2 p \cdot k\right )\right ]^{2}} =\frac{\pi  i}{4 \left (\vec{K}^{2} +2 p \cdot k\right )^{3/2}}
\end{equation}

The remaining integral is
\begin{eqnarray}\frac{\pi  i}{4} \int _{0}^{k}\frac{4 \pi ^{2} K^{2}\; d K}{\left (K^{2} +L\right )^{3/2}} &  = & i \pi ^{3} \left [\frac{ -K}{\sqrt{K^{2} +L}} +\ln  \left (K +\sqrt{K^{2} +L}\right )\right ]_{0}^{K} \\
 &  = & i \pi ^{3} \left [\frac{ -K}{\sqrt{K^{2} +L}} +\ln  \genfrac{(}{)}{}{}{K +\sqrt{K^{2} +L}}{\sqrt{L}}\right ]\end{eqnarray}

The last integral is over
$x$
. Insert
$x$
back into the integral
\begin{equation}\int _{0}^{1}\left [\frac{ -K}{\sqrt{K^{2} +2 x p \cdot k}} +\ln  \genfrac{(}{)}{}{}{K +\sqrt{K^{2} +2 x p \cdot k}}{\sqrt{2 x p \cdot k}}\right ] d x
\end{equation}

The two integrals are
\begin{equation}\int _{0}^{1}\genfrac{[}{]}{}{}{ -K}{\sqrt{K^{2} +2 x p \cdot k}} d x =\genfrac{[}{]}{}{}{ -2 K \sqrt{K^{2} +2 x p \cdot k}}{2 p \cdot k}_{0}^{1} =\frac{ -K \sqrt{K^{2} +2 m c k}}{m c k} +\frac{1}{m c}
\end{equation}

\begin{eqnarray} &  & \int _{0}^{1}\ln  \genfrac{(}{)}{}{}{K +\sqrt{K^{2} +2 x p \cdot k}}{\sqrt{2 x p \cdot k}} d x \\
 &  = & \left [\frac{k \sqrt{k^{2} +2 m c k}}{2 m c k} +\ln  \genfrac{(}{)}{}{}{\sqrt{k^{2} +2 m c k} +k}{\sqrt{2 m c k}}\right ] -\frac{k^{2}}{2 m c k}\end{eqnarray}

Substituting for
$k =\hbar  \kappa $
Eq.(\ref{vector}) and
$\eta  \equiv \hbar /2 m c r =\lambda _{C}/4\pi  r$
the mass correction is:

\begin{equation*} \Delta m \equiv \mu  \left (r\right ) =\frac{\alpha  m}{2 \pi } \left [ -\frac{\eta }{2} \sqrt{1 +\frac{1}{\eta }} +\eta  +\ln  \left \{\sqrt{\eta } \left (\sqrt{1 +\frac{1}{\eta }} +1\right )\right \}\right ]
\end{equation*}

For a stationary electron
$p =m c$
.

\end{document}